\documentstyle[12pt]{article}
\newcommand{\be}{\begin{equation}}
\newcommand{\ee}{\end{equation}}
\newcommand{\beqs}{\begin{eqnarray}}
\newcommand{\eeqs}{\end{eqnarray}}

\newcommand{\tr}{{\rm Tr}}
\begin{document}
\pagestyle{plain}
\setcounter{page}{1}
\newcounter{bean}
\baselineskip16pt

%--------+---------+---------+---------+---------+---------+---------+

\begin{titlepage}
\begin{flushright}
RU-97-75\\
UTTG-24-97\\
PUPT-1719\\
IASSNS-HEP-97/101\\
hep-th/9709091
\end{flushright}

\vspace{7 mm}

\begin{center}
{\huge Schwarzschild Black Holes}

\vspace{5mm}

{\huge from Matrix Theory }
\end{center}
\vspace{10 mm}
\begin{center}
{\large
T.~Banks\footnote
{Serin Physics Labs, Rutgers University,
Piscataway, NJ 08855.}, W.~Fischler\footnote
{Physics Department, University of Texas, Austin, TX 78712.}, 
I.~R.~Klebanov\footnote
{Joseph Henry Laboratories, Princeton University,
Princeton, New Jersey 08544.}
and L.~Susskind\footnote
{Institute for Advanced Study, Princeton, New Jersey 08540. Permanent address:
Physics Department, Stanford University, Stanford, CA 94305.}
}
\end{center}
\vspace{7mm}
\begin{center}
{\large Abstract}
\end{center}
\noindent
We consider Matrix theory compactified on $T^3$
and show that it
correctly describes the properties of Schwarzschild black holes
in $7+1$ dimensions, 
including the
energy--entropy relation, the Hawking temperature and the physical size, 
up to numerical factors of order unity.
The most economical
description involves setting the cut-off $N$ in the discretized
light-cone quantization
to be of order the black hole entropy.
A crucial ingredient necessary for
our work is the recently proposed
equation of state for $3+1$ dimensional SYM theory with
16 supercharges. We give detailed
arguments for the range of validity of this equation following
the methods of Horowitz and Polchinski.
\vspace{7mm}
\begin{flushleft}
September 1997

\end{flushleft}
\end{titlepage}

%--------+---------+---------+---------+---------+---------+---------+

\newpage
\renewcommand{\baselinestretch}{1.1} 
% include the next line for double spacing

% \renewcommand{\baselinestretch}{2}

\renewcommand{\epsilon}{\varepsilon}
\def\fixit#1{}
\def\comment#1{}
\def\equno#1{(\ref{#1})}
\def\equnos#1{(#1)}
\def\sectno#1{section~\ref{#1}}
\def\figno#1{Fig.~(\ref{#1})}
\def\D#1#2{{\partial #1 \over \partial #2}}
\def\df#1#2{{\displaystyle{#1 \over #2}}}
\def\tf#1#2{{\textstyle{#1 \over #2}}}
\def\d{{\rm d}}
\def\e{{\rm e}}
\def\i{{\rm i}}
\def\Leff{L_{\rm eff}}

%--------+---------+---------+---------+---------+---------+---------+

\def \td {\tilde }
\def \ci {\cite}
\def \sm {$\s$-model }

\def \o {\omega}
\def \inv {^{-1}}
\def \ov {\over }
\def \four{{\textstyle{1\over 4}}}
\def \fourth{{{1\over 4}}}
\def \ha {{1\ov 2}}
\def \QQ {{\cal Q}}

%To produce a box for a Dalembertian (adapted from p. 320 of TeXbook):
\def\sqr#1#2{{\vcenter{\vbox{\hrule height.#2pt
         \hbox{\vrule width.#2pt height#1pt \kern#1pt
            \vrule width.#2pt}
         \hrule height.#2pt}}}}
\def\square{\mathop{\mathchoice\sqr34\sqr34\sqr{2.1}3\sqr{1.5}3}\nolimits}

%Macros to facilitate use of halign for complicated equations:
\def\TL{\hfil$\displaystyle{##}$}
\def\TR{$\displaystyle{{}##}$\hfil}
\def\TC{\hfil$\displaystyle{##}$\hfil}
\def\TT{\hbox{##}}
%Example:
%  \eqn\One{\vcenter{\openup1\jot
%    \halign{\strut\span\TL & \span\TR & \span\TT & \span\TL & \span\TR\cr
%     x^2 &> 1 & \quad when $x$ satisfies\ \ & x &> 1 \cr
%   }}}

%List references as a continuation of the present page:
\def\shortlistrefs{\footatend\bigskip\bigskip\bigskip%
Immediate\closeout\rfile\writestoppt
\baselineskip=14pt\centerline{{\bf References}}\bigskip{\frenchspacing%
\parindent=20pt\escapechar=` Input refs.tmp\vfill\eject}\nonfrenchspacing}

%Now, some miscellany:
%Added for partial waves paper:
%\def\l{\ell}
\def\eff{{\rm eff}}
\def\abs{{\rm abs}}
\def\hc{{\rm h.c.}}
\def\+{^\dagger}

%Added for exact coefficients notes:
\def\cl{{\rm cl}}

%Added for partial waves into 3-brane notes:
\def\M{\cal M}
\def\D#1#2{{\partial #1 \over \partial #2}}

%Added for absorption by extremal 3-branes paper.  Adapted from the 
%TeXbook, p. 359.
\def\overleftrightarrow#1{\vbox{Ialign{##\crcr
     \leftrightarrow\crcr\noalign{\kern-0pt\nointerlineskip}
     $\hfil\displaystyle{#1}\hfil$\crcr}}}

%--------+---------+---------+---------+---------+---------+---------+
\def \t {\tau}
\def \td {\tilde }
\def \ci {\cite}
\def \sm {$\s$-model }

\def \o {\omega}
\def \inv {^{-1}}
\def \ov {\over }
\def \four{{\textstyle{1\over 4}}}
\def \fourth{{{1\over 4}}}
\def \ha {{1\ov 2}}
\def \QQ {{\cal Q}}

\def \lr { \lref}
\def\np {{  Nucl. Phys. }}
\def \pl {{  Phys. Lett. }}
\def \mpl {{ Mod. Phys. Lett. }}
\def \prl {{  Phys. Rev. Lett. }}
\def \pr  {{ Phys. Rev. }}
\def \ap  {{ Ann. Phys. }}
\def \cmp {{ Commun.Math.Phys. }}
\def \ijmp {{ Int. J. Mod. Phys. }}
\def \jmp {{ J. Math. Phys.}}
\def \cqg {{ Class. Quant. Grav. }}

\section{Introduction}

The problem of extreme and near-extreme black holes in string
theory has recently received a 
great deal of attention \cite{Sen,sv,juan}.
The quantum theory of D-branes \cite{dlp,polch}
and its relation to 
Supersymmetric Yang-Mills (SYM) theory \cite{EW}
has allowed for successful
qualitative (and sometimes quantitative) calculations of the
properties of these objects. 
By contrast, little has been written
about the Schwarzschild black holes in string theory, although
a rough understanding has been achieved in \cite{S93,HP}.

In this paper we take up the problem of Schwarzschild black holes in
Matrix theory \cite{BFSS}. We will see that in the particular case of
$7+1$ non-compact dimensions, enough is known about the relevant
SYM theory to derive the properties of black holes, including the
energy--entropy relation and the physical size, up to numerical
factors of order unity. In what follows, such numerical factors will
be ignored throughout the paper.

Matrix theory is best thought of as the Discretized Light-Cone
Quantization (DLCQ) of M-theory \cite{S97}, i.e. compactification
on a light-like circle of radius $R$. Accordingly, the longitudinal
momentum $P_-=P^+$ is quantized in integer multiples of
$1/R$,
\be
P_- = {N\over R}\ .
\ee
We may further compactify $d$ transverse coordinates on a $d$-dimensional
torus. For simplicity, we will often consider this torus to be
``square'' with equal circumferences given by $L$.
Another length scale that appears in the theory is the
11-dimensional Planck length, $l_{11}$.

The Matrix theory conjecture is that the sector of the theory with
given value of $N$ is exactly described by $U(N)$ SYM theory
in $d+1$ dimensions 
with 16 real supercharges. This theory lives on a dual torus with
circumferences \cite{LS,GRT}
\be \label{dual}
\Sigma\sim {l_{11}^3\over RL}
\ .
\ee
For physical applications the limit $N\rightarrow\infty$ has to be taken.
This limit is not uniform in the following sense: if we ask
how large $N$ must be taken in order to achieve a given degree
of accuracy, the answer will depend on the system under investigation.
However, choosing $N$ too large can introduce a needlessly large
number of degrees of freedom, most of which may be frozen into
their ground state. The situation is in many respects similar
to the choice of cut-off in quantum field theory, where it is
desirable to choose it so that there are neither too few nor too
many degrees of freedom. The former destroys the accuracy,
while the latter makes the calculations unnecessarily difficult.

The minimal value $N_{min}$ which will allow the desired degree
of accuracy for a black hole will certainly increase with the entropy
which, after all, is the measure of the number of relevant degrees
of freedom. Our first task will be to determine $N_{min}$.
Consider a black hole in its rest frame. The transverse
momentum $P_\perp=0$, while $P_+= P_-=M$. The transverse size of the
black hole is its Schwarzschild radius, $R_s$, and its extension
in the $X^-$ direction is also of order $R_s$. As $R_s$ grows,
it will eventually exceed the light-like compactification scale $R$,
and the black hole will not fit in the longitudinal space.
However, we may boost it, thereby Lorentz contracting it, until it does 
fit. Let us assume that it is boosted till its longitudinal
momentum is $N/R$. Its longitudinal size is then contracted
to
\be
\Delta X^- \sim {M\over P_-} R_s = {MR\over N} R_s 
\ .\ee
The condition for fitting into the transverse space is
$R> \Delta X^-$, which implies
\be\label{minimal}
N > MR_s = N_{min}
\ .\ee
Thus we see that simple kinematical considerations determine
the order of magnitude of $N_{min}$.

The Schwarzschild radius in a $D$-dimensional space-time is
\be
R_s\sim (G_D M)^{1\over D-3}
\ ,
\ee
where $G_D$ is the $D$-dimensional Newton constant.
Thus, (\ref{minimal}) becomes
\be
N_{min} \sim G_D^{1\over D-3} M^{D-2\over D-3}
\ .\ee
It is extremely interesting that the above expression is also
the Bekenstein entropy of the black hole, $N_{min} \sim S$.

Our strategy for determining the entropy--mass relation for the
black hole goes as follows. Using the Matrix theory hamiltonian $H$ for
fixed $N$ we compute the partition function, $Z=\tr e^{-\beta H}$.
From this we deduce the relation between
the energy and the entropy for given $N$,
\be
E= E(N, S)\ .
\ee
Next we observe that the Matrix theory hamiltonian is identified
with the DLCQ energy according to
\be \label{energy}
E= {M^2\over P_-}= {M^2 R\over  N}
\ .\ee
Thus, we find
\be\label{massdef}
M^2 ={N\over R} E(N,S)\ .
\ee
Now, for $N\gg N_{min}$, the value of $M^2$ computed this way must
be independent of $N$. However, as we shall see, computing the
partition function for $N\gg S$ is very difficult.
Thus, we are forced to choose $N\sim S$, and (\ref{massdef}) becomes
\be\label{newmassdef}
M^2 \sim {S\over R} E(S,S)\ .
\ee
Note that the Matrix hamiltonian is explicitly proportional to
$R$, so that $R$ cancels in (\ref{newmassdef}) leaving a relation between
mass and entropy.

\section{The case $d=3$}

One might have guessed that the easiest case to analyze is that of
$D=11$ black holes, where the matrix model is just supersymmetric
quantum mechanics. In fact, this case seems to be especially
difficult: very little is known about this system for general
$N$. The case involving the most widely studied SYM theory is
$d=3$, leading to $D=8$ black holes. Therefore, we 
concentrate on the $d=3$ case. 

The SYM theory relevant to the $D=8$ black holes is the very special
self-dual conformally invariant theory in $3+1$ dimensions with
16 real supercharges. We will begin by illustrating the strategy
outlined in the previous section without fully justifying the
formulae. More details are given in the next section.

Since the SYM theory is conformally invariant, its equation of state
must have the form
\be
\label{state} S = C \Sigma^3 T^3\ ,\qquad E = C \Sigma^3 T^4
\ ,
\ee
where $T$ is the temperature of the Matrix theory (not to be confused with
the Hawking temperature), and $\Sigma^3$ is the volume of the dual
torus. $C$ measures the number of degrees of freedom which, for
the adjoint representation of $U(N)$, is expected to be $C\sim N^2$.
This equation of state is supported by the form of the near-extremal
entropy of the self-dual 3-brane found in \cite{GKP,kt}.

Eliminating the temperature from (\ref{state}) 
and using (\ref{dual}), (\ref{energy}) gives
\be
S \sim M^{3/2} \left ( {G_{11}\over N L^3}\right )^{1/4}
\ .\ee
Now we set $N \sim S $ and use the standard expression for the
Newton constant in 8 dimensions, $G_8= G_{11}/L^3$, arriving at
\be
S\sim M^{6/5} G_8^{1/5}\ ,
\ee
which is correct for $D=8$ black holes! Note that not only does the scaling
with $M$ come out correctly, but so does the dependence on $L/l_{11}$.

Although the above derivation will prove to be correct, there
are serious questions concerning the range of validity of
(\ref{state}). At the point $S\sim N$, the temperature given by
(\ref{state}) with $C\sim N^2$ satisfies
\be\label{extra}
\Sigma T \sim 1/N^{1/3}
\ .\ee
For a conventional free field with periodic boundary conditions
the equation of state (\ref{state}) is valid only when
the temperature satisfies $\Sigma T > 1$. This is just the condition
that the wavelength of a typical thermal quantum is smaller than
the box size. Clearly, (\ref{extra}) requires us to extrapolate
the equation of state to much lower temperatures. This sort of situation
has arisen before in the theory of D-brane black holes \cite{MS,DM}
where, due to the presence of Wilson loops, the effective size
of the quantization box is much larger than its actual size.
We will return to this point in the next section and show that
this is exactly what happens when a single 3-brane is wrapped
$N^{1/3}$ times over each of the direction of the 3-torus.

Before doing this, however, let us consider implications of black hole
physics for the equation of state when $N \gg S$ or, equivalently,
$\Sigma T\ll 1/N^{1/3}$.
In this range the entropy must be 
independent of $N$ 
for given $M$. In $D=8$ the formula is $S=M^{6/5}G_8^{1/5}$. 
Using $M^2 = EN/R$ and $dE=TdS$ we find the equation of 
state
\be \label{newstate}
S=(N T \Sigma)^{3/2}    
\ .
\ee
At the point $S=N$ this agrees with our previous equation of 
state. The implication is that the eqation of state (\ref{state}) 
which holds at high temperature must continue down to 
temperature $\sim {1 \over N^{1/3}\Sigma}$ but no further. A 
transition to the equation of state (\ref{newstate}) 
must occur at this 
point. We will see in the next section that there is 
good reason to believe that just such a transition  occurs.

\section{Thermodynamics of wrapped 3-branes}

In this section we study the thermodynamics of Dirichlet 3-branes
wrapped over a rectangular 3-torus with sides of length $\Sigma_i$.
We will be particularly interested in a single 3-brane wrapped
$N_1$ times over direction $\hat 1$,
$N_2$ times over direction $\hat 2$, and
$N_3$ times over direction $\hat 3$.
The coordinates along such a 3-brane are given by
\be
X^i = {N_i \Sigma_i\over 2\pi} \theta_i\ , \qquad i=1,2,3
\ee
where $\theta_i$ are the three angles running from 0 to $2\pi$.
The total volume of such a multiply wound 3-brane is
\be
V_{tot}= N_1 N_2 N_3 \Sigma_1 \Sigma_2 \Sigma_3
\ .\ee
Therefore, its charge is the same as that of $N=N_1 N_2 N_3$ singly
wound parallel 3-branes. The dynamics of such a system is governed
by ${\cal N}=4$ supersymmetric $U(N)$ Yang-Mills theory in 3+1
dimensions \cite{EW}. To describe the multiply wound configuration, 
appropriate Wilson loops need to be introduced \cite{PCJ}.
For example, a D-string wound $N_1$ times is described in
$1+1$ dimensional SYM theory by the holonomy
which is a shift matrix: its non-zero entries are
$\phi_{i+1,i}=1$ for $i=1, \ldots, N_1-1$ and $\phi_{1,N_1}=1$.
In other words, the holonomy matrix encodes how the different
strands of the D-string are connected. Similarly, the three
$U(N)$ holonomy matrices for the multiply wound 3-brane 
encode the connections among the $N_1 N_2 N_3=N$ sheets as 
we move along the holonomy cycles.

For sufficiently large $\Sigma_i$ (or the temperature $T$) there should be
no difference between the thermodynamic properties of the multiply
wound brane and those of $N$ coincident singly wound branes.
The latter theory has $O(N^2)$ massless degrees of freedom on volume
$V_d= \Sigma_1 \Sigma_2 \Sigma_3$, 
and we find the following expressions for the
energy and the entropy,
\be \label{conf}
E\sim N^2 V_d T^4\ , \qquad S\sim N^2 V_d T^3
\ .\ee 
For the multiply wound brane the same scalings follow from a different line
of reasoning, which is based on the arguments in \cite{HP}. 
Now the fields live on the volume $V_{tot}=N V_d$,
and there are $O(N)$ massless species. The latter fact may seem surprising,
but it is a direct consequence of the D-brane theory 
\cite{dlp,polch,PCJ}. 
Indeed, the
3-brane consists of $N=N_1 N_2 N_3$ interconnected sheets, and there are 
distinct massless open
strings connecting sheet $1$ with sheet $j$, $j=1, \ldots, N$.

The difference between the two configurations is illuminated by 
T-dualizing along all three directions. The $N$ singly wound 3-branes 
are mapped into $N$ coincident 0-branes on the dual torus. The
multiply wound 3-brane is instead mapped into an array of 
0-branes \cite{Aki},
with $N_1$ rows along direction $\hat 1$,
$N_2$ rows along direction $\hat 2$, and
$N_3$ rows along direction $\hat 3$. A string connecting two 0-branes
in general has a fractional winding number along direction $i$, quantized
in units of $1/N_i$. For each allowed winding number, we find $N$ different
species because the string can start on each of the 0-branes in the array.
This implies that, before T-duality, the allowed
values of momentum in the $i$-th direction are 
quantized in units of $2\pi/(N_i \Sigma_i)$, and
we have $N$ different massless fields.

While for high enough temperature it does not matter how the $N$
3-branes are interconnected, a crucial difference appears as the temperature
is lowered. For $N$ singly wound branes, (\ref{conf}) holds approximately 
only if $T \Sigma_i > 1$. For the multiply wound brane, the momenta
$p_i$ are quantized in units of $2\pi/(N_i \Sigma_i)$, and
the condition on the temperature is much less restrictive,
\be T N_i \Sigma_i >  1 \ .
\ee
Let us assume that all three $N_i \Sigma_i$ are comparable. 
Then we find that
the lowest temperature at which (\ref{conf}) applies is 
\be
T_{crit}\sim (V_d N)^{-1/3} 
\ .
\ee
At this temperature, the entropy $S$ is of order $N$.
Thus, we can achieve adequate resolution of the black hole
($N\sim N_{min}$)
right at the edge of the range of validity of (\ref{conf}).
This fact is of crucial importance for describing the properties of
8-dimensional Schwarzschild black holes in the context of Matrix theory.

$T_{crit}$ is the temperature of the black hole in the
boosted frame.
Let us calculate the value of the Hawking temperature by
boosting $T_{crit}$ back to the rest frame of the black hole.
We find
\be T_H \sim {N\over RM} (V_d N)^{-1/3} \sim
{S^{2/3}\over M G_8^{1/3}}
\ .
\ee
Using
\be \label{msg}
M\sim S^{5/6} G_8^{-1/6}
\ ,
\ee
we have
\be \label{Ht}
T_H \sim (S G_8)^{-1/6} \sim {1\over R_s}
\ .\ee
This is indeed the expected scaling of the Hawking temperature.

Another connection of the $3+1$ dimensional
${\cal N}=4$ supersymmetric $U(N)$ Yang-Mills theory
is with the semiclassical properties of the R-R charged 3-branes
in type IIB supergravity.
This connection has been explored in considerable
detail in \cite{GKP,kt,IK,GKT,GKnew}. For example, the 
ADM energy and the Bekenstein-Hawking entropy are 
given in terms of the Hawking
temperature by relations of the form (\ref{conf}) \cite{GKP,kt}. 
For infinite 3-branes,
these relations hold down to $T=0$, but in the finite case we know
that there is a minimal temperature below which they break down.
What is the origin of such a restriction from the point of view of the
classical solution? The geometry which corresponds to the multiply wrapped
brane is
\be d s^2= f^{-1/2} (- h dt^2 + dy^i dy^i) + f^{1/2} (h^{-1} dr^2
+ r^2 d \Omega_5^2)\ ,
\ee
where
\be
f(r)= 1+ {r_3^4\over r^4}\ , \qquad h(r)= 1-{r_0^4\over r^4}\ .
\ee
We will consider the near-extremal case where $r_0\ll r_3$, and $r_3$
is related to $N$ through \cite{GKP,IK}
\be r_3^4 \sim N g (\alpha')^2
\ .
\ee
Absence of large 
corrections to the metric from the higher-derivative terms in the 
string effective action requires that $N g> 1$ \cite{IK}. 
Furthermore, for the case of finite 3-branes we will require that the
longitudinal volume at the horizon is at least of order 1
in string units, i.e.
\be
V_d {r_0^3\over r_3^3} > (\alpha')^{3/2}
\ .\ee
In terms of the temperature $T\sim r_0/r_3^2$, this condition becomes
\be
V_d T^3 > (N g)^{-3/4}
\ .\ee
Even for large $Ng$ this condition is 
more restrictive than
$V_d T^3> 1/N$ found for the multiply wound case. 
Thus, there is no contradiction between the SYM theory and the
classical calculations. The SYM approach indicates that
(\ref{conf}) does not apply for $T < T_{crit}\sim (NV_d)^{-1/3}$, 
but the classical
solution receives potentially large 
$\alpha'$ corrections starting at a higher 
temperature. Assuming agreement with SYM, we conjecture that in reality
the $\alpha'$ corrections stay small all the way down to $T\sim T_{crit}$.

\section{Estimating the size of the black hole}

Let us estimate the size of a $D=8$
black hole. As we explained, this is well described by a uniform
array of $N$ 0-branes on the original 3-torus of volume $V$.
The rms size of the black hole should
be identified with
$\sqrt{\langle \vec X^2\rangle}$ where $\vec X$ is the transverse
position of one of the 0-branes.

By the virial theorem, the 
kinetic energy of the 0-branes scales as
the total energy in the system,
\be
{M_0\over 2}\langle \sum_{i=1}^N ({d\over dt} \vec X_i)^2\rangle 
\sim N^2 V_d T^4
\ .\ee
Since $M_0=1/R$, we find that for each 0-brane
\be
\langle ({d\over dt} \vec X)^2\rangle 
\sim R N V_d T^4
\ .\ee
Using the fact that the typical frequency is of order $T$, 
we conclude that
\be \label{size}
\langle \vec X^2\rangle 
\sim R N V_d T^2
\ .\ee
Now we recall that $T_{crit}\sim (NV_d)^{-1/3}$, a
nd that $V_d$, the volume of the
dual torus, is $G_{11}/( R^3 V) $. Substituting into
(\ref{size}) we find that $R$ cancels out, as it should, and
\be \label{newsize}
\langle \vec X^2\rangle 
\sim  ( N G_{11}/V)^{1/3}
\ .\ee
Since the Newton constant in $D=8$ is
$G_8= G_{11}/V $, we finally have\footnote{
The same scaling follows if we work on the dual
torus and estimate the thermal fluctuations of scalar fields in
the SYM theory.}
\be
\sqrt{\langle \vec X^2\rangle} \sim (S G_8)^{1/6}
\ .
\ee
This is precisely the scaling with the entropy of the Schwarzschild
radius in $D=8$! It is remarkable that this scaling 
follows so simply from the Matrix theory. 

We have seen that the 0-branes spread out over the 3-torus undergo
large transverse oscillations. As a result of these oscillations, some
number of them can break off the metastable bound state and be emitted
into the transverse directions. This is the physical picture of
the Hawking radiation. Let us estimate the typical number of 
0-branes in the emitted cluster. In the rest frame of the black hole,
the typical values of $p_0$ and $p_z$ for massless Hawking particles 
are expected to be of order $1/R_s$. 
Thus, $p^{rest}_-\sim 1/R_s$ also. Now we boost this value
to the Matrix theory frame, where the total $P_-$ of the black hole
is $N/R$. In this frame the typical value for a Hawking particle is
\be
p_- \sim {N\over RM} {1\over R_s}\sim {1\over R}
\ .\ee
This means that Hawking emission proceeds a few 0-branes at a time.
Eventually the black hole completely dissociates into small clusters
of 0-branes.

\section*{Acknowledgments}

We are grateful to Steve Shenker for useful discussions.
The work of I.R.K was supported in part by the DOE grant DE-FG02-91ER40671,
the NSF Presidential Young Investigator Award PHY-9157482, and the
James S.{} McDonnell Foundation grant No.{} 91-48.  
L.S. is grateful to E. Witten for hospitality at the IAS where part of this
work was done. L.S. is a Raymond and Beverly Sackler Fellow at IAS and
is supported in part by NSF grants PHY-9219345 and PHY-9513835.
T.B. is supported in part by the DOE grant DE-FG02-96ER40559.
W.F. is grateful to Rutgers University for hospitality while
part of this work was being done.
W.F. is supported in part by the NSF under grant PHY-95-11632 
and by the Robert A.  Welch Foundation.

%--------+---------+---------+---------+---------+---------+---------+

%--------+---------+---------+---------+---------+---------+---------+

\end{document}